\documentclass[final,prl,twocolumn,superscriptaddress,showpacs,floats]{revtex4}
\usepackage{amsmath,amsfonts,amssymb,bm}
\usepackage[final]{graphicx}

\newcommand{\ex}{E^{\mbox{\tiny X}}}
\newcommand{\exx}{E^{\mbox{\tiny XX}}}
\newcommand{\exb}{E_b^{\mbox{\tiny X}}}
\newcommand{\exxb}{E_b^{\mbox{\tiny XX}}}
\newcommand{\ktroom}{k_BT_{\mbox{\tiny room}}}
\newcommand{\ryd}{Ry^*}

\begin{document}

\title{Biexciton stability in carbon nanotubes}

\author{David Kammerlander}\email{david.kammerlander@unimore.it}
\author{Deborah Prezzi}\email{prezzi.deborah@unimore.it}
\author{Guido Goldoni}
\author{Elisa Molinari}

\affiliation{CNR-INFM Research Center for nanoStructures and
bioSystems at Surfaces (S3)} \affiliation{Dipartimento di Fisica,
Universit\`a di Modena e Reggio Emilia, Via Campi 213/A, 41100
Modena, Italy}

\author{Ulrich Hohenester}
\affiliation{Institut f\"ur Physik,
  Karl--Franzens--Universit\"at Graz, Universit\"atsplatz 5,
  8010 Graz, Austria}

\date{\today}

\begin{abstract}

We have applied the quantum Monte Carlo method and tight-binding
modelling to calculate the binding energy of biexcitons in
semiconductor carbon nanotubes for a wide range of diameters and
chiralities. For typical nanotube diameters we find that  biexciton
binding energies are much larger than previously predicted from
variational methods, which easily brings the biexciton binding
energy above the room temperature threshold.

\end{abstract}



\maketitle

Size-dependent optical excitations in
nano-structures are at the heart of fundamental studies as well as
conceivable applications \cite{scholes:06}. Carbon nanotubes (CNTs)
make no exception, showing very sensitive electronic and optical
properties to the atomic structure and spanning a wide range of
wavelengths \cite{bachilo:02}.
The stability of the excitonic states, neutral or charged optically
excited electron-hole complexes, is determined by their binding
energy with respect to thermal fluctuations. In quasi-1D systems, the
binding energy can be much larger than in systems of higher
dimensionality. In inorganic semiconductor heterostructures, 
for instance, the exciton binding energy $\exb$ is
substantially larger than \cite{someya:96,ogawa:91,rossi:97}
the binding energy $4\,\ryd$ in a strictly 2D
system \cite{bastard:82}, where $\ryd$ is the Rydberg energy of the host material.
Analogously, the biexciton binding energy $\exxb$ of two electron-hole
pairs, optically excited in a two-photon process, is not
limited \cite{banyai:87,baars:98} to its 2D value of
$0.77\,\ryd$ \cite{usukura:99, hohenester:05}.

Since CNTs are quasi-1D systems, obtained by rolling up a graphene
sheet \cite{saito:98}, they are characterized by rather large binding
energies \cite{wang:05,maultzsch:05}, analogously to conjugated
polymers \cite{rohlfing:99,ruini:02}. On the other hand, in CNTs one
expects strongly diameter-dependent binding energies. Indeed, in
addition to the increase of the ratio $ E_b^{\mbox{\tiny X(XX)}} /
\ryd $ with decreasing diameter, due to the transition from a quasi-2D
system to a quasi-1D system, also electron and hole effective
masses, which determine $\ryd$, change with the CNT diameter. Furthermore, 
due to the involved energy scales, in CNTs not only excitons but also
biexcitons might be stable against thermal fluctuations at room temperature; 
also, the energy separation can be larger then the linewidth, and 
optical detection of biexcitons should be possible.
Contrary to inorganic semiconductors and semiconductor nanostructures, 
where biexcitons have received considerable interest, there is only very 
little work devoted to biexcitons in CNTs. 
Following the pioneering work of Ando \cite{ando:97}, 
the exciton binding energy has been calculated for several 
nanotubes, with different diameter and chirality, both within 
{\em ab initio} approaches~\cite{chang:04, spataru:04} and semiempirical
methods~\cite{perebeinos:04,zhao:04,pedersen:03,kane:03,capaz:06,jiang:07}.
On the contrary, the biexciton binding energy, which is presently not
accessible to the more accurate first principles methods, has been
computed only via an approximate variational approach \cite{pedersen:05}.

In this paper we use the quantum Monte Carlo (QMC) method to
calculate the \emph{exact}\/ (dimensionless) binding energy $\exxb/\ryd$
of biexcitons confined to the surface of a cylinder of 
diameter $D$. We find that the biexciton is much
more stable and its binding energy much larger than estimated from
variational methods, particularly for intermediate to large
$D/a_B^*$. Assuming homogeneous dielectric screening and
tight-binding estimates for the CNT effective masses, we also
estimate $\exxb$ for several families of CNTs. We find that
for realistic values of the dielectric constant, $\exxb$ can be
comparable or larger than $\ktroom$ even for the larger CNT diameters.

\begin{figure}
\centerline{\includegraphics[width=0.75\columnwidth,clip]{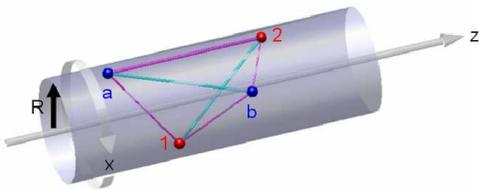}}
\caption{Biexciton complex on a cylindrical surface. $x$
is the circumferential and $z$ the longitudinal direction of the
tube with diameter $D=2R$. Particles $1,2$ (electrons) and $a,b$ (holes)
are confined to the surface and form a spin singlet state.}
\label{fig:system_sketch}
\end{figure}

Let us consider electrons and holes confined to the surface of an
infinitely long cylinder of diameter $D$, as indicated in
Fig.~\ref{fig:system_sketch}. 
It is convenient to use dimensionless exciton units, where $\hbar=1$, 
and in which masses are measured in units of the reduced electron-hole 
mass $\mu$, distances in units of the effective Bohr radius $a_B^*
=(\epsilon/\mu)a_B$, and energies in units of the effective Rydberg 
$\ryd=e^2/(2\epsilon a_B^*)$. 
With the approximation of equal electron and hole effective 
masses~\cite{pedersen:03, perebeinos:04, comment.mass}, the biexciton 
Hamiltonian in dimensionless units reads
\begin{eqnarray} \label{eq:ham}
 H&=&-\frac{1}{2}(\nabla_1^2 + \nabla_2^2 + \nabla_a^2 + \nabla_b^2 ) \nonumber\\
    &&-\frac {2}{r_{1a}}-\frac {2}{r_{1b}}-\frac {2}{r_{2a}}-\frac {2}{r_{2b}}
      +\frac {2}{r_{12}}+\frac {2}{r_{ab}}\,,
\end{eqnarray}
Here, $\bm r_1$,$\bm r_2$ ($\bm r_a$,$\bm r_b$) are the positions of
the two electrons (holes),and $r_{ij}$ is the distance between particles 
$i$ and $j$. We choose electrons (holes) with opposite spin orientations
and consider the optically active spin-singlet biexciton ground state. 

\emph{Variational QMC}.---In the following we sketch our numerical
approach. In addition to \emph{exact} QMC calculations, to be
discussed below, we have performed variational QMC
(VQMC) calculations. We have exploited the (unnormalized) Hylleraas--Ore
trial wavefunction~\cite{hylleraas:46} in a slightly modified
version
\begin{equation}\label{eq:psit}
  \Psi_T=e^{-\frac{1}{2}(s_{1a}+s_{1b}+s_{2a}+s_{2b})}
    \cosh[\frac{\beta}{2}(s_{1a}-s_{1b}+s_{2b}-s_{2a})]\,.
\end{equation}
Here, the $s_{ij}$'s are relative distances scaled by variational
parameters. As we employ cylindrical coordinates, the expression for
$s_{ij}$ reads
\begin{equation}\label{eq:scaled_dist}
s_{ij}=\sqrt{\left(\frac{D\sin{(x_{ij}/D)}}{q}\right)^2 +
\left(\frac{z_{ij}}{k}\right)^2 }\,,
\end{equation}
where $x$ and $z$ are oriented along the circumference and the
symmetry axis of the cylinder, respectively (see
Fig.~\ref{fig:system_sketch}). Variational parameters $q$ and $k$
allow for different scaling for both directions \cite{pedersen:03}.
$\beta$ is an additional variational parameter determining the
strength of the coupling between the two excitonic complexes of a
biexciton, where $\beta=1$ corresponds to two separate excitons,
and for $\beta=0$ there is equal binding within all
electron-hole pairs. The variational parameters $q$, $k$ and $\beta$
have to be determined such that the total energy
$E_T=\langle\Psi|H|\Psi\rangle/ \langle\Psi|\Psi\rangle$ becomes
minimized.

Quite generally, after separation of the biexciton center-of-mass
motion, the calculation of $E_T$ involves six-fold integrals, which
constitutes a formidable computational task. The calculation of
$E_T$ is performed by the VQMC approach \cite{thijssen:99}, whose main
elements can be summarized as follows: since the trial
wavefunction \eqref{eq:psit} of the optically active spin-singlet biexciton 
groundstate is always positive 
(thereby avoiding the fermionic sign problem), 
it can be represented by an ensemble of ``walkers'', each one characterized by
the particle positions $\bm r_1$, $\bm r_2$, $\bm r_a$, $\bm r_b$.
Starting from a suitable initial configuration, one generates a
Markov chain for the walkers where the probability for a specific
configuration is given by $\Psi_T^2(\bm r_1,\bm r_2,\bm r_a,\bm
r_b)$. Upon sampling of the ``local energy'' $E_L=H\Psi_T/\Psi_T$
one then obtains the energy $E_T$ associated to the trial
wavefunction \cite{thijssen:99}. Let us denote the ensemble of
walkers with $\rho(\bm r_1,\bm r_2,\bm r_a,\bm r_b, t)$, where $t$
is a fictitious time. The Fokker-Planck equation, which in our
dimensionless units reads
\begin{equation}\label{eq:fp}
  \frac{\partial\rho}{\partial t}=\frac 14\sum_{i=1,2,a,b}{\nabla_i
  \left(\nabla_i-\bm F_i\right)}{}\rho,
\end{equation}
in time-discretized form defines a scheme to proceed from a
configuration $\rho(t)$ to $\rho(t+\delta t)$ according to the drift
and diffusion process given on the right-hand side \cite{comment.drift}.
Thus, the VQMC simulation consists of the three main
steps of (i) initialization of the ensemble of walkers, (ii) drift
and diffusion of all particles in each walker according to
Eq.~\eqref{eq:fp}, and (iii) sampling of the local energy $E_L$ once
the stationary distribution is reached.

\emph{Guide Function QMC}.---A slight variant of the VQMC approach
allows for the exact solution of the Schr\"odinger equation. Let
$\rho=\Psi\overline{\Psi}_T$ denote a function composed of the exact
wavefunction $\Psi$ and the \emph{guide function}
$\overline{\Psi}_T$. The Fokker-Planck-type
equation
\begin{equation}\label{eq:qmc}
  \frac{\partial\rho}{\partial t}=\frac 14\sum_{i=1,2,a,b}{\nabla_i
  \left(\nabla_i-\bm F_i\right)}{}\rho-\left[E_L-E\right]\rho
\end{equation}
in time-discretized form again defines a scheme that can be solved
by means of Monte-Carlo sampling. It can be easily
proven~\cite{thijssen:99} that under stationary conditions
Eq.~\eqref{eq:qmc} reduces to the exact Schr\"odiger
equation. Thus, once the invariant $\rho$ is obtained the exact wavefunction is at hand.
We represent $\rho$ by an ensemble of walkers, and account on the
right-hand side of Eq.~\eqref{eq:qmc} for the first term through
drift and diffusion and for the second term through a Monte-Carlo
branching with probability $p=\exp[-(E_L-E)\delta t]$.
Depending on the value of $p$, the walker dies, survives, or gives 
birth to other walkers \cite{thijssen:99}.
In the simulation
the energy $E$ is chosen such that the total number of walkers
remains approximately constant and a constant distribution $\rho$ is
reached. This invariant distribution and $E$ then
determine the biexciton wavefunction and energy, respectively.
Therefore, the main steps of this so-called \emph{guide function}
QMC approach are (i) initialization of the ensemble of walkers, (ii)
drift and diffusion of all particles in each walker, (iii) branching
of the walkers, and (iv) sampling of the wavefunction once the
stationary distribution is reached.

Technically, one needs to choose $\delta t$ sufficiently small to
allow for the separate drift-diffusion and branching steps
accounting for the two different terms on the right-hand side of
Eq.~\eqref{eq:qmc} \cite{tech-note}; $\overline{\Psi}_T$ has to be 
chosen such
that the local energy $E_L$ in Eq.~\eqref{eq:qmc} remains finite
when two particles in a walker approach each other. While this is
guaranteed for the true wavefunction, $\overline{\Psi}_T$ is usually
taken as a Jastrow-type wavefunction with the correct cusp
condition \cite{jastrow:55,thijssen:99}. In practice we use
$\overline{\Psi}_T = \Psi_T $ with $q=k=\frac 1 4$, $\beta=0$.
We finally emphasize that both the VQMC and the exact QMC
simulations can be applied in a straightforward manner to excitons,
in which case the trial wavefunction is of the form $\Psi_T(\bm
r_1,\bm r_a)=\exp[-s_{1a}]$ and $\overline{\Psi}_T =\exp[-2
r_{1a}]$, respectively.

\begin{figure}[t]
\centerline{\includegraphics[width=0.95\columnwidth,clip]{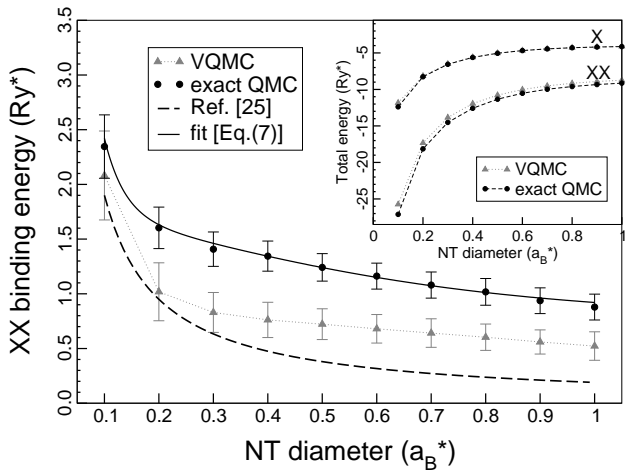}}
\caption{Biexciton binding energies as a function of CNT diameter, calculated
with exact guide function QMC (black dots) and VQMC method (grey triangles).
Statistical error bars of both methods are shown. 
The solid, black line is the result of using our fitting functions in 
Eq.~\eqref{eq:fitting}. Inset: total energy of exciton and biexciton. 
Note that here $\ex =
-\exb$.  }\label{fig:binding}
\end{figure}

\begin{figure}
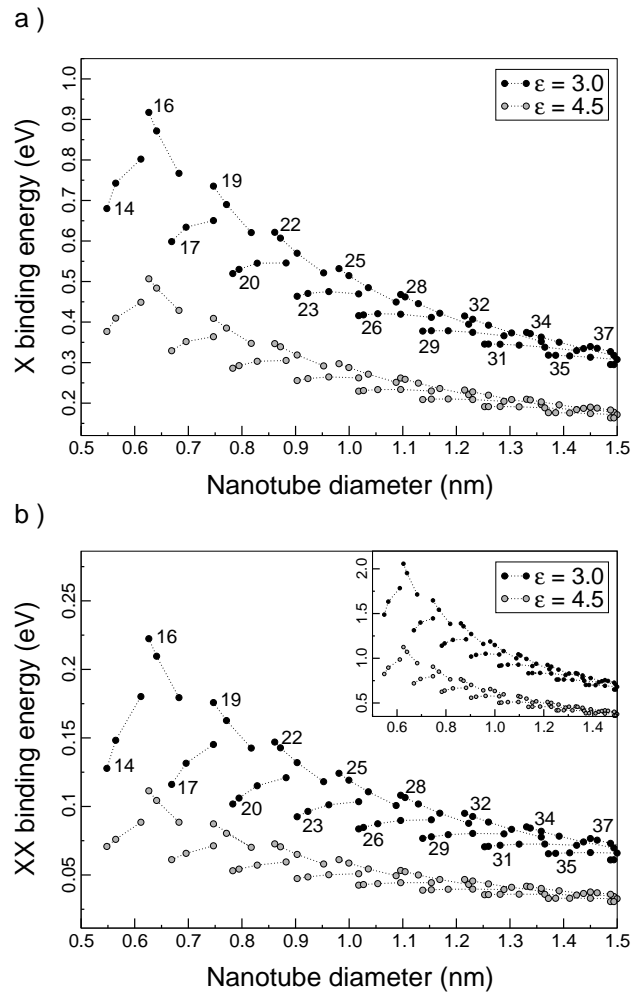

\includegraphics[width=0.95\columnwidth,clip]{exciton.eps}
\vfil
\includegraphics[width=0.95\columnwidth,clip]{biexciton.eps}
\caption{Exciton (a) and biexciton (b) binding energies as a function of CNT diameter
and chirality. Inset: biexciton energies ($-\exx$) in absolute units. 
The labels indicate the (2n+m) families, where (n,m) are the chiral 
indices of the NT.}\label{fig:kataura}
\end{figure}

\emph{Results}.---In the inset of Fig.~\ref{fig:binding} we show the
dimensionless total energies $\ex$ and $\exx$ with respect to the band gap ($E_g=0$),
calculated exactly by the guide function QMC method. As can be
seen, the exciton energy $\ex$ shows the correct behaviour at infinite
diameter, that is the 2D limit, where it converges to 
$-4\, \ryd$~\cite{bastard:82}, while
it is strongly red-shifted as the diameter is decreased due to the
larger binding of the electron-hole pair, showing the transition
from a quasi-2D to a quasi-1D system. Analogously, the biexciton
energy $\exx$ red-shifts as a result of the increased interaction of
the two electron-hole pairs. 

In order to investigate the stability
of the biexciton complex with respect to the formation of separated
excitons, we show in Fig.~\ref{fig:binding} the exact biexciton
binding energy $\exxb$ for arbitrary dimensionless diameter $D/a^*_B$,
along with the VQMC results and the fitting function of 
Ref.~\onlinecite{pedersen:05}.
The biexciton results to be stable ($\exxb>0$) at any
diameter, and it shows the expected limiting behaviour at
infinite diameter, that is the 2D limit $0.77 \, \ryd$~\cite{usukura:99}. 
The binding energy increases with decreasing diameter, showing again 
the transition from a quasi-2D to a quasi-1D system. 
As can be noted, both variational results severely underestimate 
the binding energy in the whole range of diameters except for very small 
values. 
Moreover, they do not show the correct 2D limiting behavior. Such shortcoming 
of Hylleraas-Ore-type wavefunctions is in agreement with corresponding 
calculations for two-dimensional quantum wells \cite{heller:97,usukura:99}.
To give a rough estimate, for typical CNT diameters of
$D = 0.8 \div 1.2$ nm and using the dielectric constant $\epsilon = 3.5$
given in Ref.~\onlinecite{pedersen:05} the excitonic units are 
$a^*_B = 2.5 \div 5.5$ nm and $\ryd  = 0.04 \div 0.08$ eV. This brings
$\exxb$ into the $0.06 \div 0.12$ eV range, which is $1.5 \div 2.5$ larger 
than the variational results reported in Fig.~\ref{fig:binding}.

In order to calculate exciton ($\mbox{\scriptsize X}$) and 
biexciton ($\mbox{\scriptsize XX}$) absolute energies explicitly, we assume 
that excitonic effects ($\mbox{\scriptsize X}$ or $\mbox{\scriptsize XX}$ 
binding) can be decoupled from band structure effects. We can therefore 
write the energies as
\begin{equation}\label{eq:decoupling}
E^\chi = f^\chi(D/a_B^*) \ryd(D)\,,
\end{equation}
where $\chi\in\{\mbox{\scriptsize X},\mbox{\scriptsize XX}\}$ and $
f^\chi(x)$ is the exact dimensionless excitonic or biexcitonic
energy shown in the inset of Fig.~\ref{fig:binding}. 
We provide below a fitting function for $ f^\chi(x)$, which allows 
us to calculate absolute binding energies for an arbitrary diameter,
\begin{equation}\label{eq:fitting}
f^\chi (x) =
\left( a_\chi x^{-1} + b_\chi  x^{-2} +  c_\chi x^{-3} \right) \exp
(-d_\chi x) + f^\chi_{\mbox{\tiny 2D}}\,,
\end{equation}
where $f^\chi_{\mbox{\tiny 2D}}$ is the correct 2D limit,
and $a_\chi, b_\chi, c_\chi, d_\chi$ are the
fitting parameters summarized in table I. The
quality of the fit is proven by calculating the biexciton binding energy
with $2 f^{\mbox{\tiny X}}-f^{\mbox{\tiny XX}} $, see
solid black line in Fig.~\ref{fig:binding}.

We calculate the Rydberg energy $\ryd$ from the tight-binding model 
of Ref.~\onlinecite{jorio:05}, which provides explicit fitting functions 
[Eq.~(2) of Ref.~\onlinecite{jorio:05}] for the electron and hole effective 
masses of semiconducting CNTs of arbitrary chirality and diameter. 
For the dielectric screenig entering $\ryd$, we adopt the simplest 
screening model, in which all Coulomb interactions are reduced to an
effective static dielectric constant $\epsilon$, following current 
literature~\cite{pedersen:03,perebeinos:04,zhao:04,maultzsch:05}. 
This approximation, which proved to be successful in comparison with 
both experiments~\cite{wang:06} and ab-initio calculations~\cite{capaz:06}, is 
particularly suitable for embedding media with large dielectric constant, 
i.e. $\epsilon \gtrsim 3$, where the dielectric response is dominated by 
that of the medium rather then by the CNT polarizability~\cite{jiang:07}.

Figure~\ref{fig:kataura} shows Kataura-like plots for both
exciton (a) and biexciton (b) binding energies in absolute units,
in the $0.5 \div 1.5$-nm diameter range, calculated from
Eq.~(\ref{eq:decoupling}) and for different value of $\epsilon$. 
As expected, both $\mbox{\scriptsize X}$ and $\mbox{\scriptsize XX}$ 
energies decrease with increasing tube diameter, as follows from the 
behaviour of $ f^\chi(x)$. The chirality effects, entering through the 
effective masses, introduce a modulation in the binding energy dependence 
on the diameter, significantly spreading out the binding energies for the 
range of diameters considered. As shown in Fig.~\ref{fig:kataura}(b),
however, the biexciton binding energy is predicted to be above the
$\ktroom$ threshold (26 meV)
even for the largest CNTs.

In summary, we have performed QMC calculations of the singlet
optically active biexciton binding energy for CNTs of arbitrary
diameter and chirality. The biexciton has been found to be always stable
at room temperature in typical dielectric environments. We have also
developed a scheme which allows us to calculate the exact exciton and
biexciton binding energies through a simple fitting function
approach, which includes the strong correlation effects exactly and
the band-structure effects within a tight-binding approach.

\begin{table}[t!]
\caption{Fitting parameters in Eq.~\eqref{eq:fitting} for exciton (X) and
biexciton (XX) energies in dimensionless exciton units.
          }\label{tab:fitting}
\begin{ruledtabular}
\begin{tabular}{cccccc}
    & $a_\chi$ & $b_\chi$ & $c_\chi$ & $d_\chi$ & $f^\chi_{\mbox{\tiny 2D}}$ \\
\hline
  $\mbox{\scriptsize X}$ & -2.62 & 0.3024 & -0.01504 & 2.795 & -4.00 \\
  $\mbox{\scriptsize XX}$ & -5.08 & 0.56 & -0.02832 & 2.345 & -8.77 \\
\end{tabular}
\end{ruledtabular}
\end{table}


\begin{thebibliography}{99}
\bibitem{comment.mass}
We performed additional QMC calculations for different electron and hole masses (mass difference of ten percent) and found that the resulting changes of the biexciton binding energy fall well within the error bars given in Fig.~\ref{fig:binding}.
\bibitem{comment.drift}
In the drift and diffusion step, each
particle in a walker is displaced by the ``force'' $\bm F_i$ and finally a random displacement, which in dimensionless
units reads $\eta\sqrt{\delta t/2}$, is added, where $\eta$ is drawn
from a Gaussian distribution with variance one. It can be easily
checked in Eq.~\eqref{eq:fp} that for the force term $\bm
F_i=2\nabla_i\Psi_T/\Psi_T$ the invariant distribution $\rho$, i.e.,
the distribution that does not change with time, is given by
$\Psi_T^2$.
\bibitem{tech-note}
We use typical values of $10^4$ walkers, $\delta t=10^{-3}$, and
$3\times 10^4$ Monte-Carlo steps, where the first $10^4$ steps are
used to reach equilibrium. We checked that our results are not
noticeably modified for smaller values of $\delta t$.
\end{thebibliography}
\end{document}